\documentclass[%
 aip,
rsi,%
 amsmath,
 amssymb,
 reprint,%
]{revtex4-1}

\usepackage{graphicx}
\usepackage{dcolumn}
\usepackage{bm}

\begin{document}


\title{A compact radiofrequency drive based on interdependent resonant circuits for precise control of ion traps}


\author{Amelia Detti}
\email{detti@lens.unifi.it}
\affiliation{Universit\`a degli Studi di Firenze, Dipartimento di Fisica e Astronomia, via G. Sansone 1, \\I-50019 Sesto Fiorentino, Italy}
\affiliation{Istituto Nazionale di Ricerca Metrologica, Strada delle Cacce 91, I-10135 Torino, Italy}
\author{Marco De Pas}
\affiliation{Universit\`a degli Studi di Firenze, Dipartimento di Fisica e Astronomia, via G. Sansone 1, \\I-50019 Sesto Fiorentino, Italy}
\affiliation{European Laboratory for Nonlinear Spectroscopy, Via N. Carrara 1, I-50019 Sesto Fiorentino, Italy}
\author{Lucia Duca}
\affiliation{Istituto Nazionale di Ricerca Metrologica, Strada delle Cacce 91, I-10135 Torino, Italy}
\author{Elia Perego}
\affiliation{Istituto Nazionale di Ricerca Metrologica, Strada delle Cacce 91, I-10135 Torino, Italy}
\affiliation{Politecnico di Torino, corso Duca degli Abruzzi 24, I-10129 Torino, Italy}
\author{Carlo Sias}
\affiliation{Istituto Nazionale di Ricerca Metrologica, Strada delle Cacce 91, I-10135 Torino, Italy}
\affiliation{European Laboratory for Nonlinear Spectroscopy, Via N. Carrara 1, I-50019 Sesto Fiorentino, Italy}
\affiliation{INO-CNR, via N. Carrara 1, I-50019 Sesto Fiorentino, Italy}
\date{\today}

\begin{abstract}
Paul traps are widely used to confine electrically charged particles like atomic and molecular ions by using an intense radiofrequency (RF) field, typically obtained by a voltage drop on capacitative electrodes placed in vacuum. We present a RF drive realized on a compact printed circuit board (PCB) and providing a high-voltage RF signal to a quadrupole Paul trap. The circuit is formed by four interdependent resonant circuits --- each of which connected to an electrode of a Paul trap --- fed by low-noise amplifiers, leading to an output voltage of peak-to-peak amplitude up to 200\,V at $3.23$\,MHz. The presence of a single resonant circuit for each electrode ensures a strong control on the voltage drop on each electrode, e.g. by applying a DC field through a bias tee. Additionally, the moderate quality factor $Q=67$ of the resonant circuits ensures a fast operation of the drive, which can be turned on and off in less than $10$\,$\mu$s. Finally, the RF lines are equipped with pick-ups that sample the RF in phase and amplitude, thus providing a signal that can be used to actively control the voltage drop at the trap's electrodes. Thanks to its features, this drive is particularly suited for experiments in which high trap stability and excellent micromotion compensation are required.  
\end{abstract}

\maketitle
Paul traps are one of the most exceptional tools in the field of experimental quantum physics\cite{Paul:1990}. By applying an RF signal to one or more electrodes, it is possible to create a deep attractive potential capable of trapping ideally anything possessing an electric charge, from macroscopic objects\cite{Winter:1991} to atomic and molecular ions\cite{Leibfried:2003}. Charged particles can be trapped for a considerably long time, especially with the use of laser cooling techniques, making crystals of cold ions confined in a Paul trap one of the most promising hardware for the realization of quantum technologies\cite{Haeffner:2008}. 
In its linear configuration, a Paul trap is formed by two pairs of parallel, linear electrodes (e.g. rods) that are equally spaced. Opposite electrodes are placed at the same voltage in order to create an electric quadrupole field at the center of the trap. Trapping is achieved when one pair of electrodes is connected to ground, while the second one is placed at a voltage $V_0$\,sin$(\Omega_T\,t)$.
In an alternative configuration, a pair of opposite electrodes is fed with a voltage $V_0/2$\,sin$(\Omega_T\,t)$, while the second pair is placed at $V_0/2$\,sin$(\Omega_T\,t+\pi)$. Both configurations lead to an attractive harmonic pseudo-potential of frequency\cite{Leibfried:2003}: 
\begin{equation}
\omega_r=\frac{q}{\sqrt{2}m}\frac{V_0}{r_0^2\Omega_T},
\end{equation}
where $q$ is the particle's charge, $m$ its mass, $r_0$ the distance between the center of the trap and the electrodes, and $\omega_r\ll\Omega_T$.
For most applications in ion trapping, a trapping frequency $\omega_r > 1$\,MHz is highly desirable, since it facilitates the optical control of the ion motional state in the trap\cite{Monroe:1995}. To this end, a relatively large amplitude of the RF potential is needed. In a non-miniaturized trap, for instance, a typical distance between the electrodes ranges from 100s\,$\mu$m to a few mm, and the amplitude of the RF needed must range from a few hundreds to thousands of volts. For a given amplitude of RF, a method to maximize the frequency of the secular motion is to feed all the electrodes with an RF signal having the right phase. 

One of the main challenges in realizing a Paul trap is finding a way to feed the electrodes with such a large oscillating voltage. 
A common strategy is to realize a resonant circuit in which the ion trap --- playing the role of the capacitance --- is connected in series to an inductor\cite{Jones:2000}. This approach has two advantages: on the one hand there is no need of high power, since the amplification is provided by the resonance of the resulting LC circuit; on the other hand, the resonant circuit plays the role of a filter, thus reducing the effects of the noise associated to the primary source. In order to maximize both effects, the quality factor $Q$ of the resonator should be as large as possible. To this end, the resonant circuit must have a large inductance and a low resistance, while the capacitance should be ideally provided only by the ion trap, in order to maximize the voltage drop across the rods.

 In a typical implementation, the inductance is provided by a helical resonator, formed by a few windings of copper wire surrounded by a conducting shield\cite{Macalpine:1959,Siverns:2012}. 
However, this device usually has a relatively large size, which poses difficulties in the design of a compact experimental setup. 
As a result, increasing interest has been devoted to the miniaturization of RF drives into printed circuit boards (PCBs)\cite{Mathur:2006}, for instance by using logic CMOS inverters\cite{Jau:2011} or low noise amplifiers\cite{Noriega:2016}.

In this paper we present an RF drive realized on a PCB and formed by a network of four resonant circuits --- each connected to a single electrode of a linear Paul trap --- in which the high voltage output is obtained with the combination of a resonant RLC circuit and a low-noise commercial amplifier. The four resonant circuits are interconnected in order to ensure the existence of a common resonant mode with the same frequency and amplitude. This design, which differs from other RF drives recently realized on PCBs\cite{Mathur:2006, Jau:2011,Noriega:2016}, ensures that an independent DC field can be provided to each of the four RF outputs with a bias tee.
This condition is particularly important in applications in which an excellent compensation of micromotion is needed\cite{Berkeland:1998}, e.g. in atom-ion physics experiments\cite{Sias2014}. Finally, the RF drive is provided with two different kinds of pick-up probes, which continuously measure both the amplitude and the phase of the RF. These signals can be used to monitor the voltage generated by the drive and to actively stabilize the potential acting on the ions \cite{Johnson:2016}.  
Each resonant circuit has a quality factor $Q\sim67$, but the combination with a low power amplifier ensures an overall increase of the voltage by a factor $\sim 200$. The moderately small $Q$ of each resonant circuit ensures a fast operation of the drive, a property that can be relevant in experiments in which the trapping potential has to be quickly turned off, e.g. for transferring a particle from a Paul trap to an optical trap\cite{Schneider:2010}.

The paper is organized as follows: in the first paragraph we present the conceptual design of the interdependent resonant circuits. In the second paragraph we show the RF board, while in the third paragraph we present some measurements of characterization of the board. Finally, the last paragraph is devoted to the conclusions.

\section{Resonant circuit design}

The RF drive is formed by four resonant circuits --- one per electrode of a Paul trap --- that amplify and filter the RF signal. The resonant circuits are interconnected in the RF drive in order to enforce a common resonating frequency. In this section we provide a detailed description of the characteristics of each resonant circuit, and of how they are combined in the RF board. 

%

\subsection{The single resonating circuit}

The fundamental circuit of the RF drive is an RLC circuit resonating at the frequency $\Omega_T= 
\sqrt{1/L_{\text{eq}} C_{\text{eq}}}$, where $L_{\text{eq}}$ and $C_{\text{eq}}$ are the equivalent inductance and capacitance of the whole circuit, respectively. 
The quality factor of this circuit is:
\begin{equation}
Q = \frac{1}{R}\sqrt{\frac{L_{\text{eq}}}{C_{\text{eq}}}}~,
\end{equation}
with $R$ being the equivalent series resistance. If we imagine to feed this circuit with a signal of frequency $\Omega_T$ and amplitude $V_0/2$, for instance through a transformer, the voltage drop at the capacitor of the RLC circuit is $V_0/2 \sin(\Omega_T t)$. The circuit is designed such that the capacitive load $C_{\text{eq}}$ is mainly due to the capacitance of the RF electrode. Its value is considered to be a fixed quantity, since it depends on the Paul trap geometry, i.e. the electrode's shape, size and the distance from the other electrically conductive parts.
The equivalent inductance $L_{\text{eq}}$, instead, is a free parameter that can be tuned in order to find the best compromize between a large quality factor $Q$ and a relatively large resonant frequency $\Omega_T$. Additionally, and independently from the resonant frequency, the quality factor is increased by reducing the total resistance $R$, which should be ideally kept as low as possible.

\begin{figure}[th]
\centering
\includegraphics[width=0.48\textwidth]{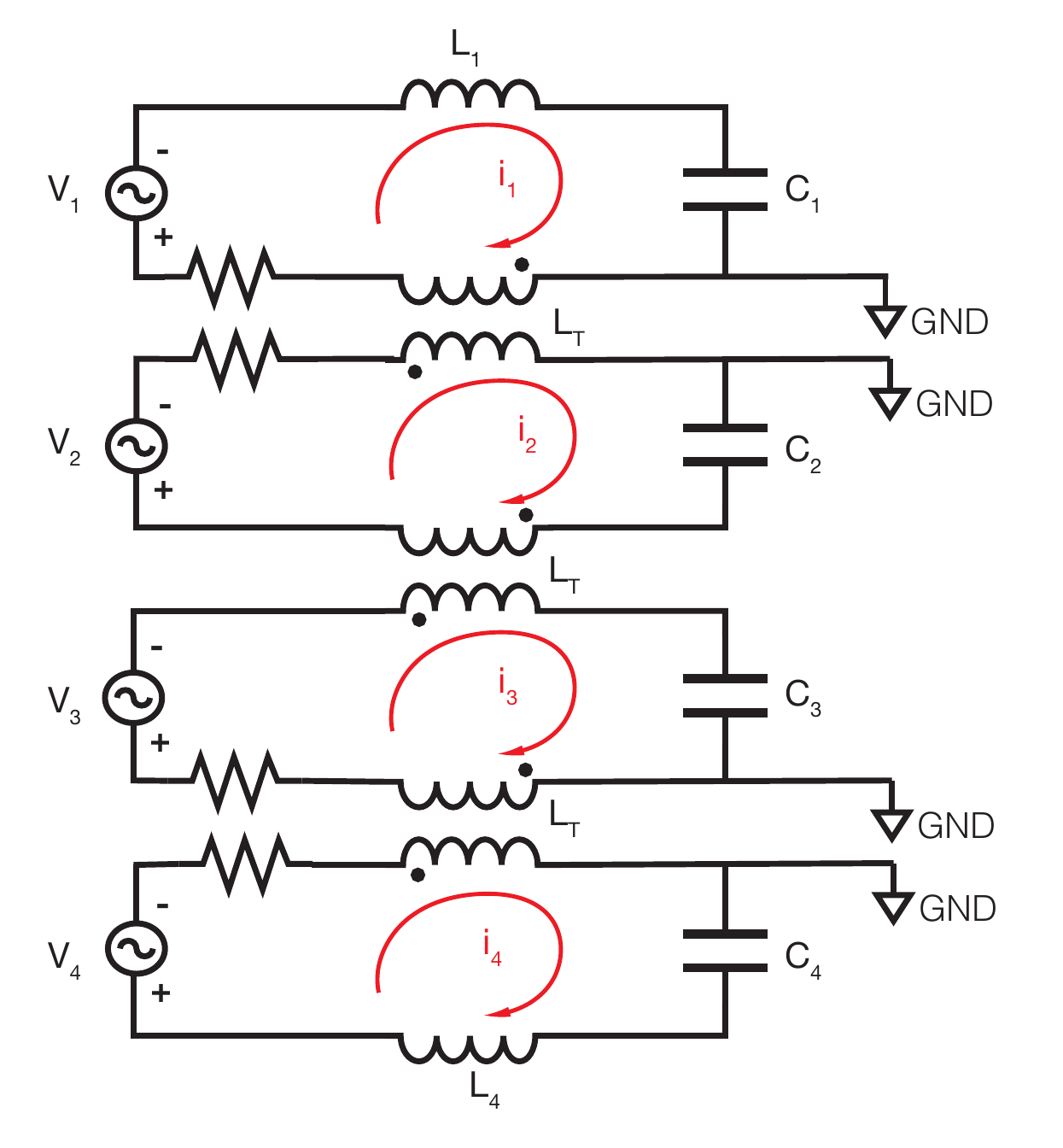}
\caption{Schematic of the RF drive representing the conceptual idea behind the design. Four resonant RLC circuits feed four capacitors - representing the total capacitance of the ion trap plus the drive circuit - at an iso-frequency $\Omega_T$.}
\label{fig:FourMeshes}
\end{figure}

A possible strategy for realizing a sufficiently large inductance -- e.g. on the order of a few 100s $\mu$H --- without implementing a bulky helical resonator is to build inductors by winding a conducting wire on a ferrite core. Ferrite cores are characterized by a frequency-dependent magnetic permeability that depends on the material, the size and the shape of the core. While the reactive term of the impendance is proportional to the real part of the magnetic permeability, its imaginary part provides a resistive term, physically caused by the  eddy currents induced in the core. In general, the resistive and reactive parts are linked by the dissipation factor DF, a specific characteristic of the core material. The presence of the inductors' resistive term affects the overall resistance $R$ of the RLC circuit, which then reads: 
\begin{equation}R=R_0 + \Omega_T L_{\text{eq}} \text{DF},\end{equation} 
where $R_0$ is the component of $R$ that does not depend on the drive frequency, originating for example from the wires' resistance. Other possible frequency-dependent terms of the resistance, for instance due to the skin effect on the wires, have for the moment been neglected. 
The quality factor of the RLC circuit can then be re-written in the form:
\begin{equation}
\label{eq:Qlimit}
Q = \frac{1}{R_0  \Omega_T C_{\text{eq}} + DF}~.
\end{equation}
In the low resistance regime ($R_0 \Omega_T C_{\text{eq}} \ll \text{DF}$), the real limitation to the $Q$ factor is the dissipation factor DF: in order to keep a high value of $Q$ it is crucial to choose a ferrite with the lowest possible losses.

%


The passive RLC circuit must be fed with RF produced by a primary source, typically a synthesizer. The connection between the primary source and the RLC resonator, or ``secondary circuit", can be realized by using a transformer. Intuitively, one may imagine that --- in order to transfer as much power as possible --- the secondary circuit must be impedance-matched to the primary one (which has typically an impedance $Z_S=50$\,$\Omega$). 
This  implies  that  at  resonance,  i.e. when the reactive part of the RLC circuit is zero, the real part of the inductance should be matched to Zs with an additional circuit connected to the primary source.
However, if $\text{Re}[Z_S]$ is smaller than $R$, impedance matching causes a decrease in the quality factor, since $\text{Re}[Z_S] + R<2R$.

As a result, in the case $\text{Re}[Z_S]<R$ it is more convenient not to match the impedances of the primary and secondary circuits, since there is not any particular constraint on the current circulating in the drive but there is the need to have a high Q and a low dissipation on the electrodes. The drawbacks for the mismatch are a higher dissipated power on the source -- a synthesizer with a high-power low-impedance amplifier--- rather than on the trap, and the possible presence of reflections creating stationary waves on the resonant lines, resulting in a DC offset on the RF electrode. However, this effect can be counter-acted by adding to the RF signal an external DC field.

\subsection{The interdependent four resonant circuits}
\begin{figure}[tbh]
\centering
\includegraphics[width=0.49\textwidth]{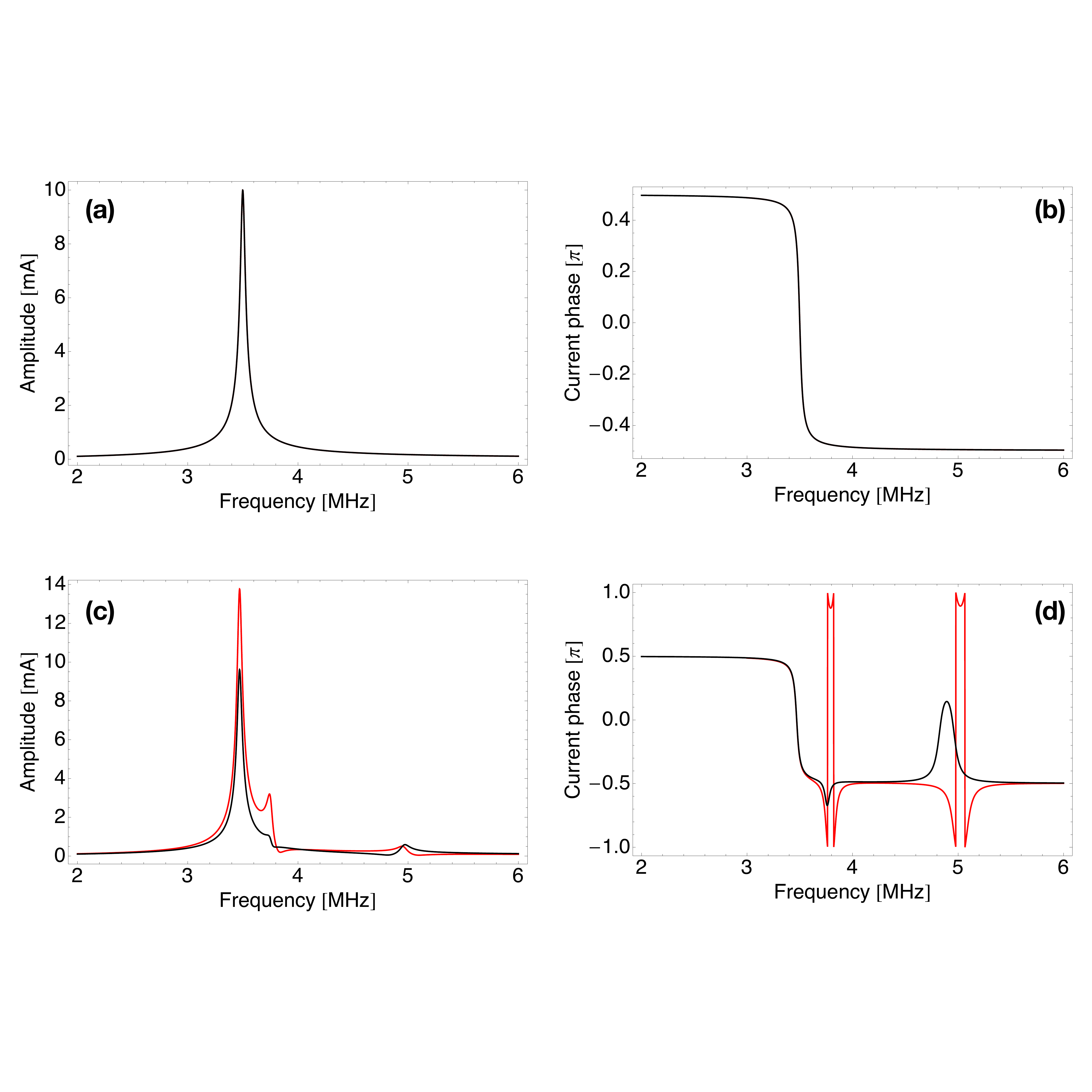}
 \caption{Simulated amplitude and phase of the current circulating in the circuit as a function of the frequency. Plots (a) and (b) show the current's amplitude and phase in the four resonant circuits in which all the capacitances are equal to $C_{\text{eq}}=5.6$~pF, all the resistances are set to $100~\Omega$, and $L1=L2=L_{\text{eq}}/2$, $L_T = L_{\text{eq}}/4$, $L_{\text{eq}}=370~\mu$H, with $L_T$ being the transformers' mutual inductances. Plots (c) and (d) show amplitude and phase of the current $i_1$ (red line) and $i_2$ (black line) in a circuit in which the electric components have the same values as in the plots (a), (b), except for $C_1=1.1~C_{\text{eq}}$ and \mbox{$C_2=0.9~C_{\text{eq}}$.}\label{fig:balance}}%
       \end{figure}

One electrode of a Paul trap can be modelled as a capacitive load. However, it is reasonable to associate to it also an equivalent series resistance (ESR), which may originate e.g. from the dielectric losses of the electrode's insulating support. Ideally, capacitance and resistance should be the same for each electrode; in practice, this is never exactly true due to possible asimmetries in the trap assembly or to mechanical imperfections. This makes it practically very hard to realize four independent RLC circuits having the same resonant frequency. Our strategy to ensure the presence of a common resonant frequency is to place $1:1$ transformers that connect the four resonant circuits (see Fig.~\ref{fig:FourMeshes}). We use a commercial software (Mathematica) to simulate this configuration, in which we consider only the resonant RLC circuits connected to an ideal RF source. The results of the simulation are shown in Fig.~\ref{fig:balance}. If one assumes that the electrodes are exactly equal, there is only one common resonance frequency $\Omega_T$, as expected. Furthermore, the RF signal in neighbour circuits has opposite phase, so the board generates in overall two pairs of identical RF signals with opposite phase, as requested for a correct functioning of the Paul trap. In case a small difference between the capacitances of the electrodes is introduced, a number of minor resonances appear, but the common resonant frequency holds, proving that this circuit design is robust against small variations of the electrical components.

%


\section{The RF drive prototype}
\begin{figure*}[t]
\centering
\includegraphics[width=0.9\textwidth]{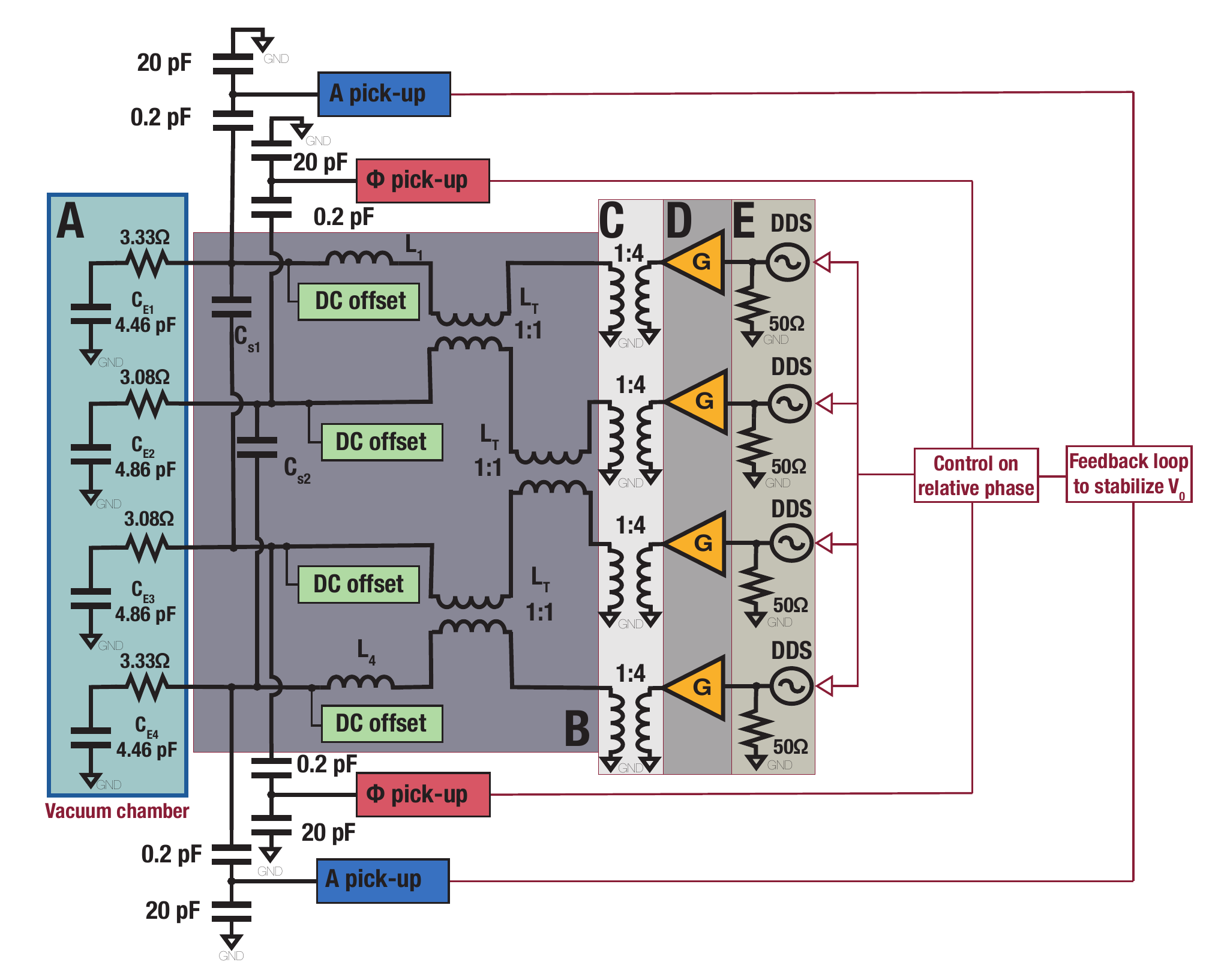}
\caption{Simplified block diagram of the complete RF drive. \textbf{A} Electrodes of a Paul trap, considered as a series of a capacitance and a resistance. The values in figure are obtained by simulating with COMSOL Multiphysics the Paul trap for which this RF drive was designed. \textbf{B} Iso-frequency resonant lines with pick-ups for phase and amplitude stabilization, balancing capacitors $C_{S1}$ and $C_{S2}$ and DC offset inputs. \textbf{C} $1:4$ step-up transformers that inductively couple the resonant circuits to the RF sources. \textbf{D} Low-impedance op-amps for pre-amplification of the input signals. \textbf{E} Four independent DDS sources providing the RF signals with tunable amplitude and phase. 
}
\label{fig:schematic}
\end{figure*}

\begin{figure}[htb]
\centering
\includegraphics[width=0.47\textwidth]{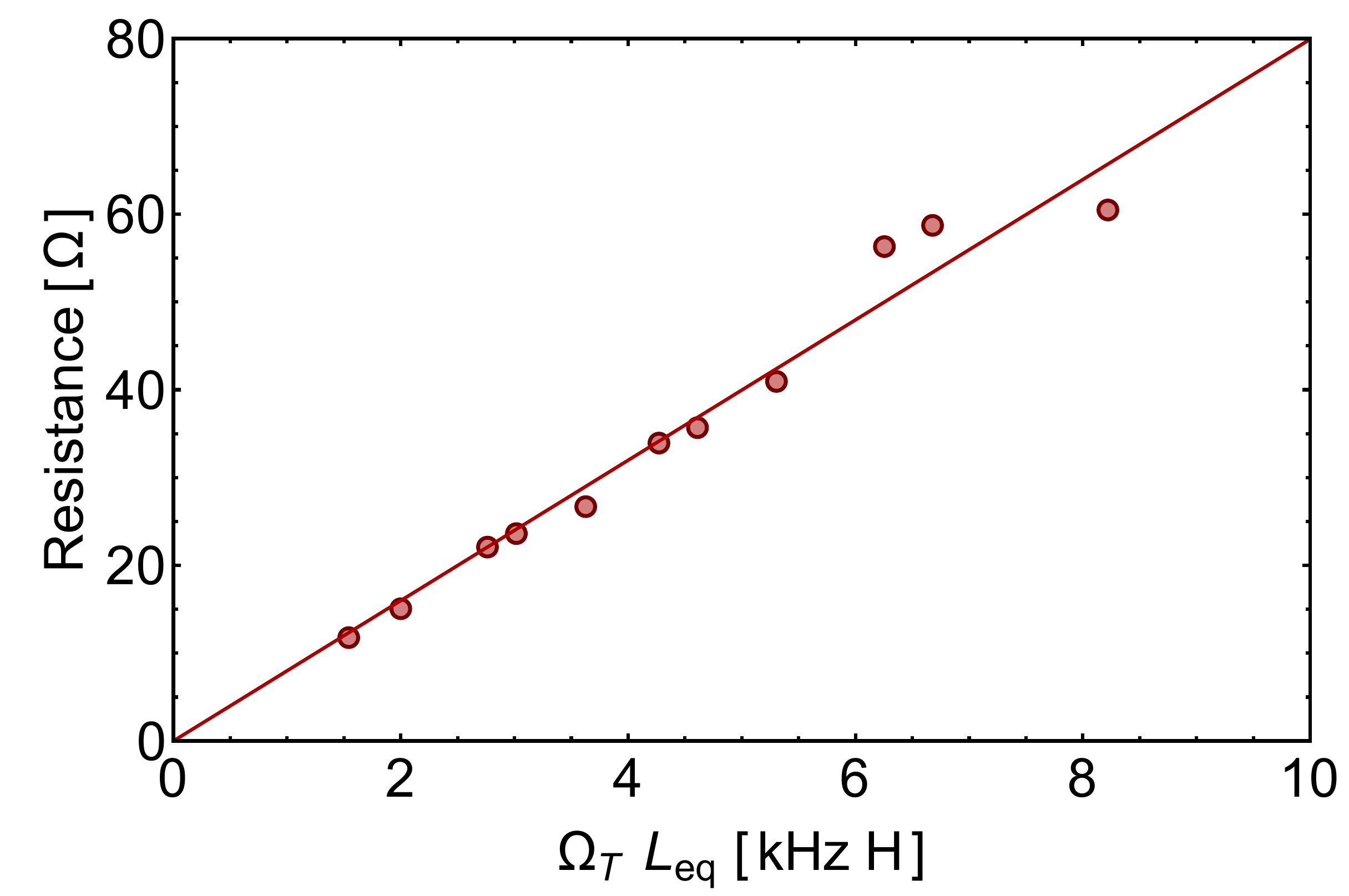}
\caption{Calibration of the dissipation factor DF for DN5H ferrite cores of height $6$~mm, internal radius $10$~mm and external radius $18$~mm. The copper wire used had a diameter of $0.2$~mm. Each point in the plot corresponds to the resistance at resonance of an RLC circuit formed by a ceramic capacitor and a homemade inductor. The values of resistance, inductance and resonant frequency are obtained with a vector network analyzer. The dependence of the total resistance as a function of $L_{\text{eq}}$ and $\Omega_T$ is extracted. The value \mbox{DF$= 0.0080\pm0.0002$} is extracted from the linear fit to the data.
}
\label{fig:DF}
\end{figure}


\begin{figure}[htb]
\centering
\includegraphics[width=0.4\textwidth]{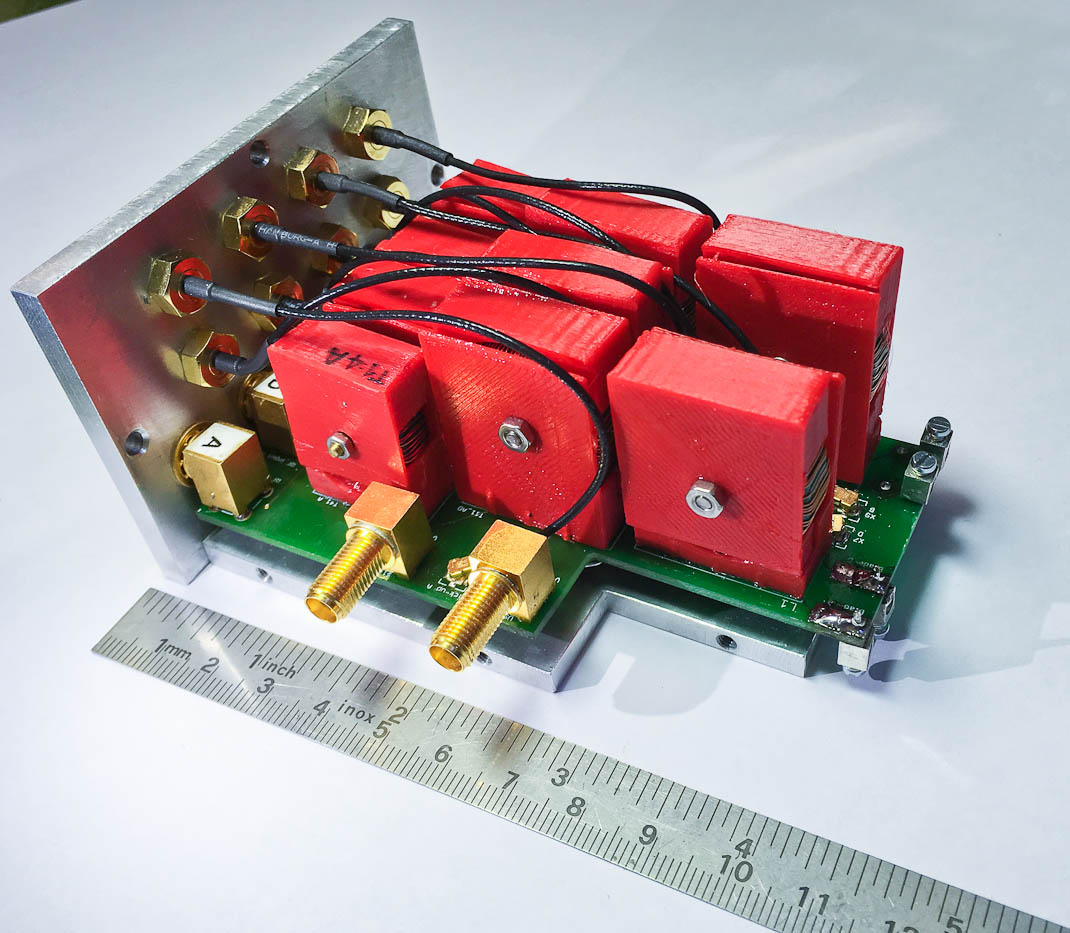}
\caption{Picture of the RF drive implemented on a 4-layer PCB. The home made inductors are incapsulated in small plastic cases (in red) in order to avoid capacitative couplings with the PCB conducting planes.}
\label{fig:pic}
\end{figure}

We have realized an RF drive based on the scheme presented in the previous section.
The scheme is divided in five main ``blocks" (see Fig.~\ref{fig:schematic}). The block \textbf{A} represents the part of the circuit that is placed inside an ultra-high vacuum chamber, i.e.\ the capacitances and the ESRs of the Paul trap electrodes. The actual values of the capacitances and the ESRs there reported are relative to the Paul trap for which this RF drive was initially designed, and they were estimated by simulating the trap with a finite element software (Comsol Multiphysics). The block \textbf{B} represents the portion of the secondary circuits that is placed outside the vacuum chamber: the blocks \textbf{A} and \textbf{B} constitute four interdependent RLC circuits. The inductances in block \textbf{B} are designed to resonate at approximately $\Omega_T =2\pi\times3.5$~MHz. Block \textbf{C} is formed by four $1:4$ transformers that are used to connect the primary and the secondary circuits, i.e. to feed the resonant lines with RF signal. The purpose of these step-up transformers is to increase by a factor $a=4$ the voltage amplitude from the primary to the secondary circuit, thereby enhancing the total amplification factor.

Conceptually, the RF sources of Fig.~\ref{fig:FourMeshes} are realized by the blocks \textbf{D} and \textbf{E}. Block \textbf{E} is constituted by four independent Direct Digital Synthesis (DDS) chips that are set at a frequency $\Omega_T/2\pi$ and have tunable amplitudes and phases. They are connected to four amplifiers (block \textbf{D} --- Analog Devices AD8392) characterized by an output impedance of $R_S = 0.2~\Omega$ at $\Omega_T =2\pi\times3.5$~MHz.The amplifiers' input impedance is matched to the $50~\Omega$ output of the DDSs.

%

The op-amps are set for a feedback gain of $G\simeq5$. This choice ensures an increase of the signal enhancement at resonance, since the real part of the impedance of the circuit is reduced. If we consider the pre-amplification stage and the step-up transformers, the $Q$ factor of Eq.~(\ref{eq:Qlimit}) becomes
\begin{equation}\label{eq:Qtot}
Q = \frac{1}{(R_0 + a^2 R_S) \Omega_T C_{\text{eq}} +\text{ DF} }~.
\end{equation}

The magnetic cores used for the inductors of the RF drive are toroidal Ni-Zn ferrite cores (material: DN5H from the company DMEGC), which are characterized by a low dissipation factor around $\Omega_T$. In order to estimate the cores' dissipation factor, we have realized a number of resonant circuits formed by a single inductor and a conventional capacitor resonating at approximately $\Omega_T$. With these circuits, we have measured the resistance at resonance as a function of the impedance of the inductor. The results are plotted in Fig.~\ref{fig:DF}. A dispersion factor DF$= (8.0\pm0.2)\times10^{-3}$ is obtained from a linear fit to the data. By substituting this value in Eq.\,(\ref{eq:Qtot}), we can estimate $Q\simeq 115$, and a resonance frequency of $\Omega_T =2\pi\times 3.5$~MHz. We notice that the dispersion factor DF is the largest term in the denominator of Eq.\,(\ref{eq:Qtot}), and therefore the dispersion of the core represents the strongest limitation to the $Q$ factor of this RF drive.

Since the capacitance and the ESR of the electrodes are in general not identical, we have placed two balancing capacitors $C_{S1}$ and $C_{S2}$ that connect the in-phase lines of the drive in order to ensure that the RF signals of the secondary circuits have all the same amplitude. 
Moreover, we have equipped the RF drive with phase and amplitude pick-ups that can be used to perform active stabilization of the RF signal, with the goal of counter-acting possible effects --- like thermal fluctuations of the different elements of the drive --- that may affect the long-term stability of the trap frequency $\omega_r$ ~\cite{Johnson:2016}.
%
%
%
%
%
%
The active stabilization of the RF potential requires a faithful sampling of the voltage drop at the electrodes. This needs to be done in a section of the circuit that is at a high impedance load. To this end, we implemented little-invasive probes with $1:100$ capacitive dividers realized with the $0.2$~pF and $20$~pF capacitors as the last elements of the circuit outside the vacuum (see Fig.\ \ref{fig:schematic}). 
The capacitive dividers work as phase pick-ups ($\phi$-pick-ups) that carry information on the resonance frequency $\Omega_T$ and the relative phase between the signals on the RF electrodes. Therefore this signal can be used both to stabilize the trap frequency $\omega_r$ by changing $\Omega_T$, and to actively stabilize the relative phase between the electrodes.
The amplitude pick-ups (A-pick-ups) are realized by placing a rectifier after the capacitive divider to generate a signal linearly dependent on the RF amplitude $V_0/2$. The board has one A-pick-up and one $\phi$-pick-up for both pairs of electrodes at $V_0/2$\,sin$(\Omega_T\,t)$ and at  $V_0/2$\,sin$(\Omega_T\,t+\pi)$.


\section{Characterization of the RF drive}
We have realized the RF drive on a 4-layers PCB of size approximately 100\,mm$\times$70\,mm (see Fig.\ \ref{fig:pic}).
The operation of the board was first characterized by substituting the electrodes with $5.6$~pF ceramic capacitors. This choice allowed us to directly monitor the RF signal with external probes, if needed. 

\begin{figure}[tbh]
\centering
\includegraphics[width=0.48\textwidth]{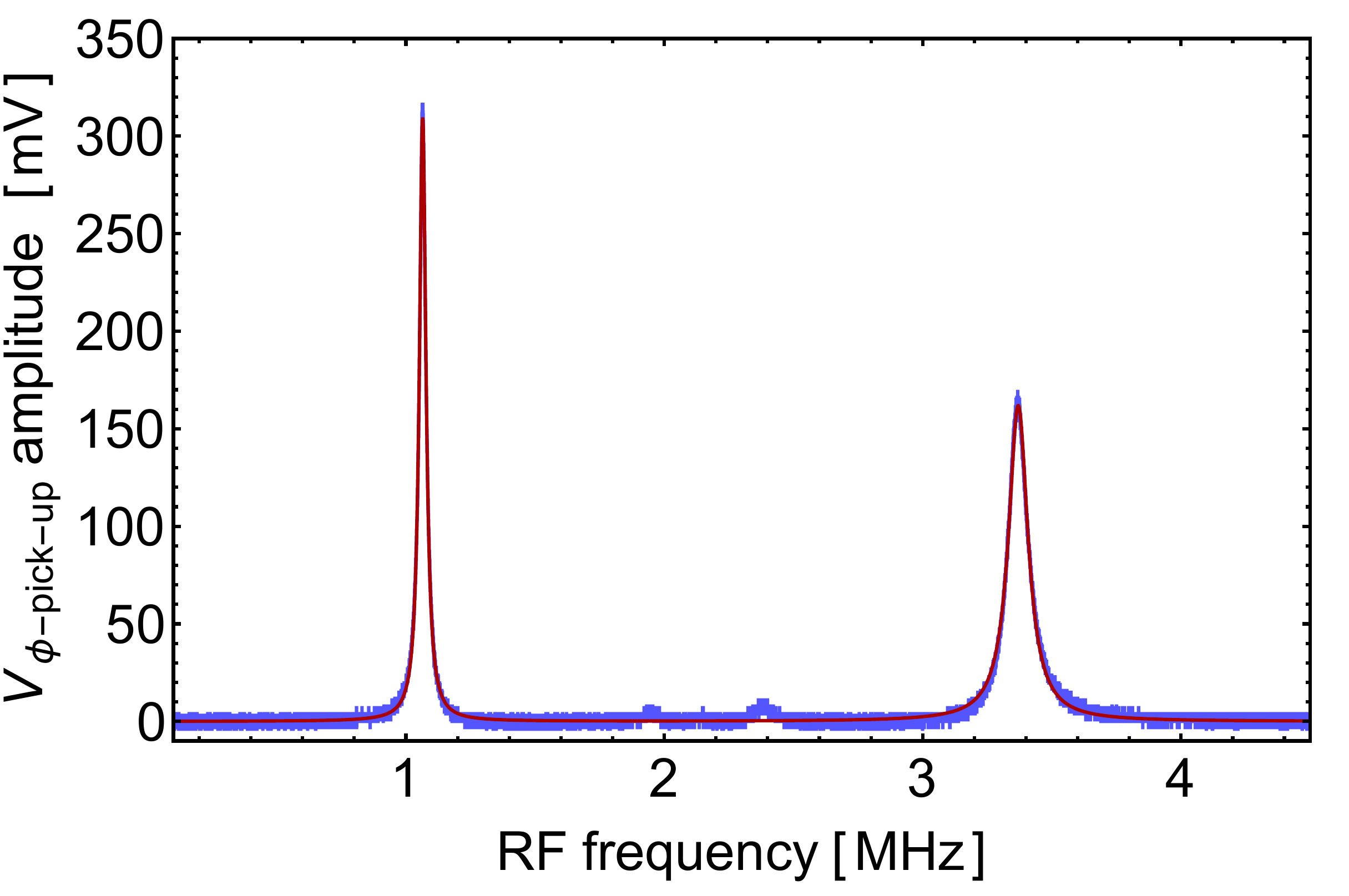}
\caption{Plot of the drive response tested with $5.6$~pF ceramic capacitors emulating the trap electrodes. The frequency of the input RF signals is swept to observe the main resonances of the circuit. The peak-to-peak amplitude of the RF signal reaching $C_{E2}$ is detected via the $\phi$-pick-up output and fitted with a Lorentzian curve (red) to extract the linewidth. Among all the resonances, the one at 3.37~MHz is the only one for which the RF signals of the four secondary circuits have the right phase relation.}
\label{fig:resonance}
\end{figure}

Fig.\ \ref{fig:resonance} shows the amplitude of the RF signal in the secondary circuit as a function of the frequency of the primary RF source. The circuit has several resonances, including the one at $\Omega_T =2\pi\times3.37$~MHz that we had initially targeted as the working resonance, and that we expected from the simulation to be at approximately $2\pi\times3.5$~MHz. The resonance at $\Omega_T $ has a full width at half maximum of $2\pi\times 91$~kHz, which we extracted from a Lorentzian fit. The resulting quality factor is $Q= 57$, which is lower than what expected from Eq.\,(\ref{eq:Qtot}).
We attribute this discrepancy to the non-ideal behavior of the transformers that have a mutual inductance coefficient smaller than $1$. Other effects that may reduce the quality factor of the circuit are possible cross-talks between transformers that are closely placed on the board. We notice that the value of the quality factor could be increased e.g.\ by changing the inductors of the board, at the price of lowering the resonant frequency $\Omega_T$. The relatively small quality factor ensures a fast operation of the drive: we measure a falling time at resonance of $7.7$\,$\mu$s, corresponding to a time constant $\tau = 3.5$\,$\mu$s.

\begin{figure}[tbh]
\centering
\includegraphics[width=0.48\textwidth]{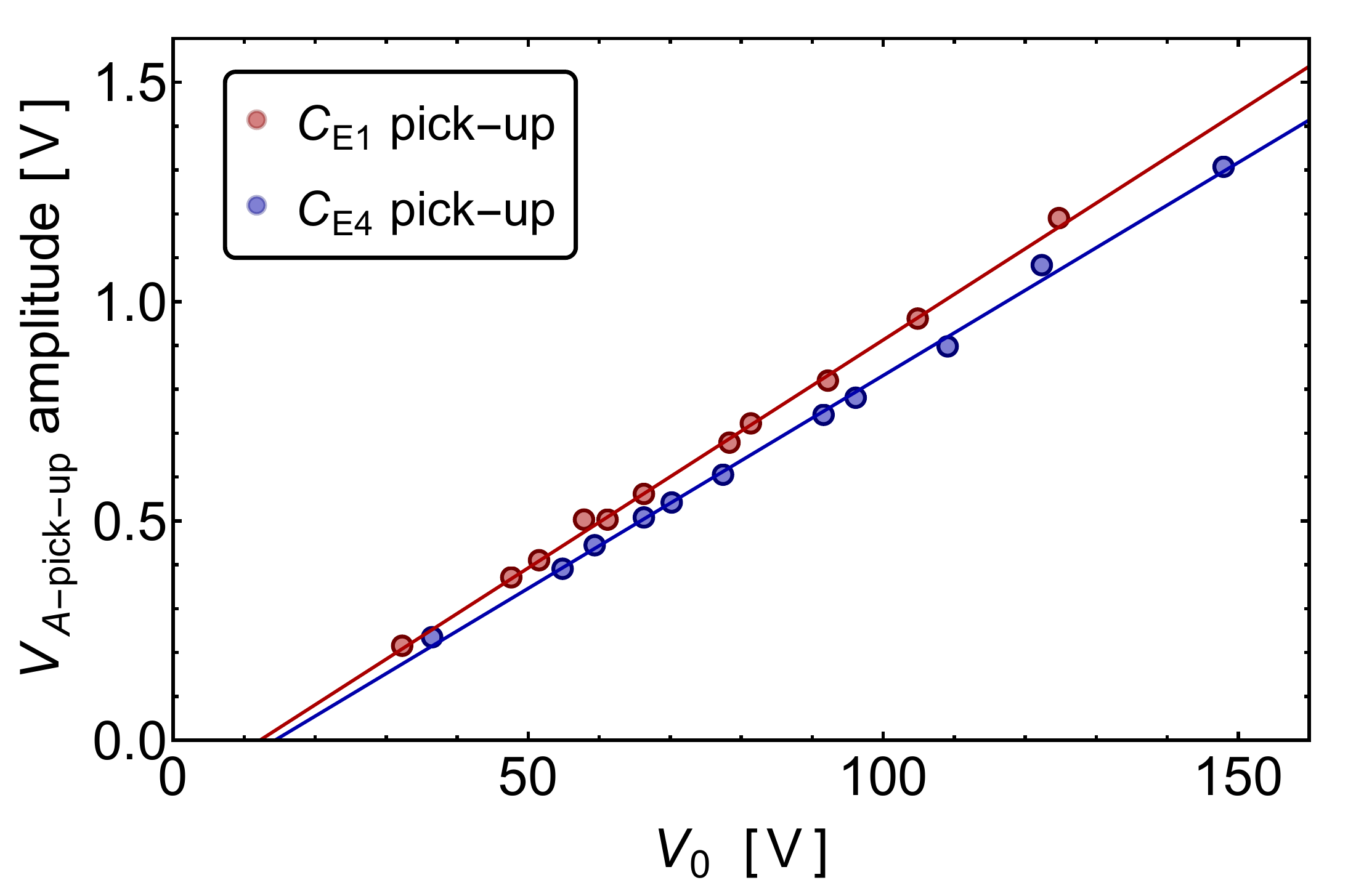}
\caption{Linearity test for the two A-pick-ups sensing the signals reaching $C_{E1}$ and $C_{E4}$. The measurement is performed by changing the amplitude of the RF signal, $V_0/2$. Experimental data are plotted together with their linear fits, which lead to the slopes $10.4\pm0.2$ and $9.7\pm0.2$ for the pick-ups placed close to $C_{E1}$ and $C_{E4}$, respectively. The difference between the two curves are due to small mismatches in the values of the electric components forming the capacitive dividers and the rectifiers. The linear fitting functions have a non-zero offset caused by the non-linear behaviour at low RF amplitudes of the diodes that are part of the rectifiers.}
\label{fig:pickupLin}
\end{figure}

\begin{figure}[tbh]
\centering
\includegraphics[width=0.48\textwidth]{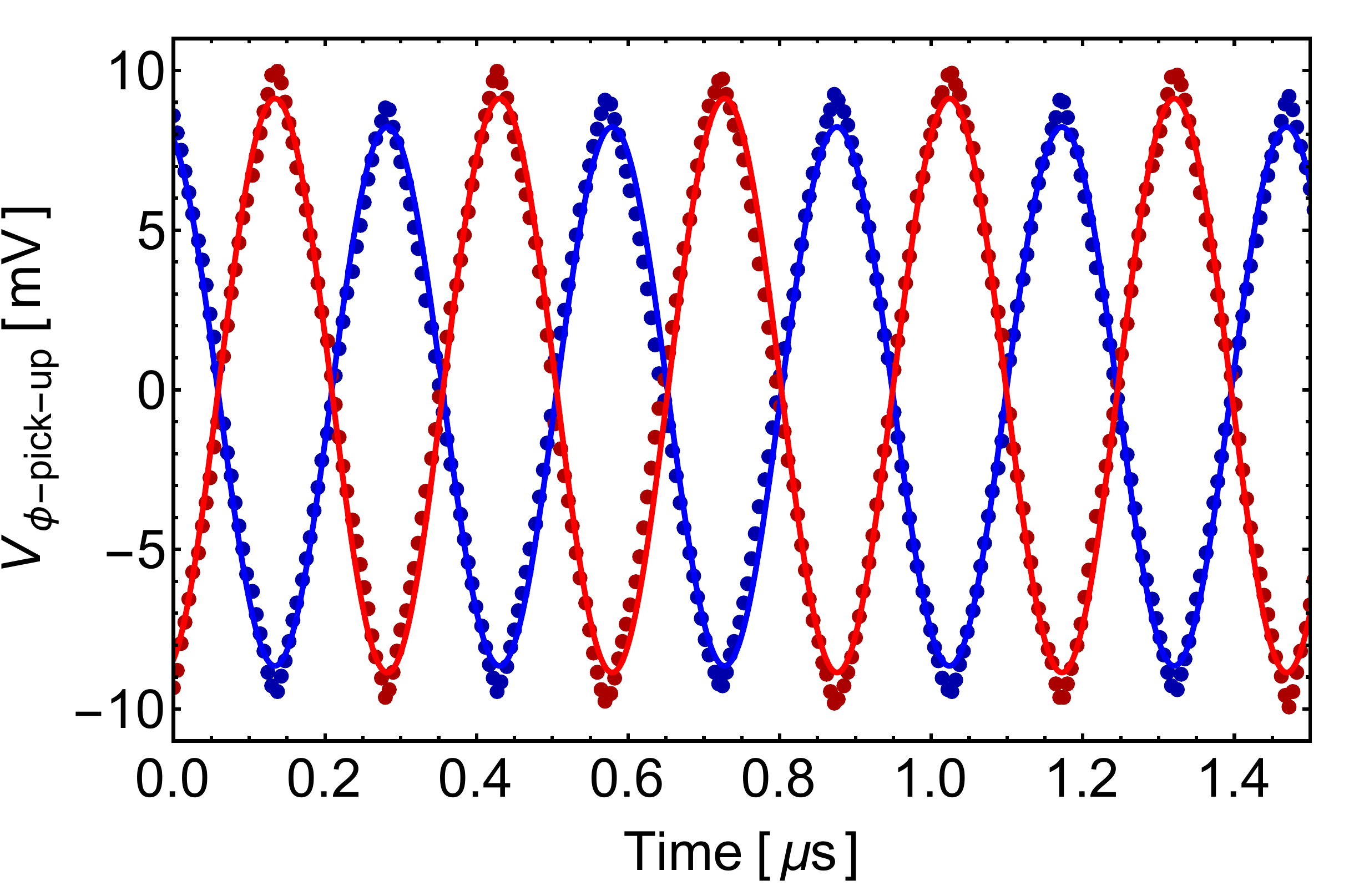}
\caption{Phase of the signal outputs.  
The data show the outputs of the $\phi$-pick-ups probing the signals at the capacitors $C_{E2}$ (red) and $C_{E3}$ (blue). Measurements are taken with the 
DDSs frequencies set to $\Omega_T$. When the phase of the input signals are set correctly, the outputs of the two $\phi$-pick-ups are sinusoidal signals that are perfectly out of phase. The sinusoidal fit to the data gives a phase difference $\phi/\pi=1.001 \pm 0.006$. }
\label{fig:AC}
\end{figure}

In addition to the resonance at $\Omega_T$, other resonances arise at different frequencies (see Fig.\ \ref{fig:resonance}). We attribute the presence of these resonances to the small unbalancing between the four resonant circuits, the components of which are not perfectly equal. This speculation was confirmed by simulating the circuit with the commercial software LTSpice, in which a small variation of one of the components of the circuit results in the appearance of new resonances.
Of all the additional resonances, we further investigated the strongest one at about $1.06$~MHz. This resonance corresponds to the creation of four RF signals having all the same phase. This condition clarifies the appearance of this resonance at a frequency considerably smaller than $\Omega_T$: since the equivalent inductances associated to the 1:1 transformers in the block \textbf{B} of Fig.~\ref{fig:schematic} depend on the direction of the currents, a resonance with all the RF signals in phase is characterized by values of $L_{eq}$ that differ with respect to the ones associated with the resonance at $\Omega_T$, resulting in a strong difference in both the frequency and the quality factor of the resonance. 
%

We tested the response of the pick-ups at resonance. Fig.\ \ref{fig:pickupLin} shows the response of the two A-pick-ups as a function of the amplitude of the primary RF source. The data show a very good linear response. Additionally, we measured the bandwidth of the A-pick-ups, finding it equal to $6$~Hz. This corresponds to the maximum bandwidth for a stabilization loop that can correct slow amplitude variations. If needed, this bandwidth can be changed by modifying the components of the rectifiers. This can be done at the cost of increasing the ripple amplitude on the pick-up signal --- currently approximately $15$~mV for $V_0=200~V$. Additionally, we used the $\phi$-pick-ups to measure the relative phase between the RF signals used to feed neighbour electrodes. For an optimal operation of the Paul trap, this phase should be $\pi$. Fig.\,\ref{fig:AC} shows the output signals of the two $\phi$-pick-ups when phase frequency and amplitude of the DDSs are set correctly. The sinusoidal signals of frequency $\Omega_T$ have a phase difference $\phi/\pi=1.001 \pm 0.006$, measured from fit.  We note that the amplitude of the A-pick-ups and the $\phi$-pick-ups are slightly different, possibly because of small differences in the components forming the capacitive dividers. These differences can be easily calibrated by using external probes.

At resonance, we are able to produce signals of peak-to-peak amplitude up to $V_0=200$~V with an input RF of peak-to-peak amplitude $1$~V. We checked the stability of the drive by running it for 24 hours at the maximum output without active stabilization. We found no appreciable changes of the resonant frequency, but we noticed a small variation of the phase of the RF signal. These small instabilities, possibly due to thermal effects on the ferrite core, can be actively corrected by probing the signals with the $\phi$-pick-ups. The total power consumption of the board is due to the op-amps and is $\sim 2$~W. This value, which is calculated for the sum of all four resonant circuits, currently limits the maximum reachable peak-to-peak voltage, since a higher power could lead to a failure of the op-amps. We note that the power consumption is nominally higher than other compact drive implementations in literature. However, the power consumption in these drives is kept low by realizing smaller gains\cite{Jau:2011}, by running at lower frequencies \cite{Noriega:2016}, or by working at temperatures for which a cryostat is needed \cite{Gandolfi:2012}.

Finally, we tested the RF drive when connected to the linear Paul trap that is currently under development at our laboratory. We measure a resonance at $\Omega_T=2\pi\times 3.23$~MHz, with a full width at half maximum of $2\pi\times 75$~kHz, resulting in a quality factor $Q=67$. This small variation originates from the fact that in its actual implementation we found the electrodes to have a capacitance of approximately $10$\,pF, slightly larger than what initially simulated.

\section{Conclusions}

We presented a compact RF drive for ion traps built on a PCB and composed of four interdependent resonant RLC circuits which are capable to work in iso-frequency, iso-amplitude and proper phase relation. The presence of four resonant circuits, one per electrode of a Paul trap, ensures the possibility of adding to each RF signal a DC voltage through a bias tee. A high voltage output is realized by using a low-noise amplifier and the enhancement factor of the resonant RLC circuits. The drive has a quality factor $Q=67$  --- mainly limited by the the dissipation factor of the ferrite material constituting the inductors of the resonant circuits --- while the total voltage gain of the whole drive is approximately 200. This gain is sufficiently large to ensure ion trapping while dissipating moderate powers on the op-amps, namely less than 2~W for four trap electrodes. The drive is also equipped with non-invasive pick-ups in order to monitor the RF signals amplitude and phase, making it possible to actively stabilize the voltage drop on the electrodes of the Paul trap via feedback loops.

Quantitatively, the performances of the drive in its current realization depend on the specific ion trap to which it is connected. However, the drive schematics can be easily adapted to other traps with different capacitive loads and different trap frequencies, possibly leading to better performances. Additionally, the drive scheme can be scaled up to a larger number of inter-connected resonant circuits, e.g. for driving multipoles RF traps\cite{Wester:2009}. Finally, the compact PCB size and low power consumption make the drive suitable for transportable ion trap experiments.
\\
\\


 \begin{acknowledgments}
We thank Massimo Inguscio for continuous support, the members of the LENS electronic workshop for discussions and Claudio Calosso for discussions and a critical reading of the manuscript. This work was supported by the ERC-Starting Grant PlusOne (grant agreement no. 639242), the SIR grant ULTRACOLDPLUS (grant code RBSI14GNS2), the FARE grant UltraCrystals (grant code R165JHRWR3), and the the European Metrology Programme for Innovation and Research (EMPIR) project 17FUN07 (CC4C). This project has received funding from the EMPIR programme co-financed by the Participating States and from the European Union’s Horizon 2020 research and innovation programme.
 \end{acknowledgments}

\bibliography{RFdrive_bib}

\begin{thebibliography}{17}%
\makeatletter
\providecommand \@ifxundefined [1]{%
 \@ifx{#1\undefined}
}%
\providecommand \@ifnum [1]{%
 \ifnum #1\expandafter \@firstoftwo
 \else \expandafter \@secondoftwo
 \fi
}%
\providecommand \@ifx [1]{%
 \ifx #1\expandafter \@firstoftwo
 \else \expandafter \@secondoftwo
 \fi
}%
\providecommand \natexlab [1]{#1}%
\providecommand \enquote  [1]{``#1''}%
\providecommand \bibnamefont  [1]{#1}%
\providecommand \bibfnamefont [1]{#1}%
\providecommand \citenamefont [1]{#1}%
\providecommand \href@noop [0]{\@secondoftwo}%
\providecommand \href [0]{\begingroup \@sanitize@url \@href}%
\providecommand \@href[1]{\@@startlink{#1}\@@href}%
\providecommand \@@href[1]{\endgroup#1\@@endlink}%
\providecommand \@sanitize@url [0]{\catcode `\\12\catcode `\$12\catcode
  `\&12\catcode `\#12\catcode `\^12\catcode `\_12\catcode `\%12\relax}%
\providecommand \@@startlink[1]{}%
\providecommand \@@endlink[0]{}%
\providecommand \url  [0]{\begingroup\@sanitize@url \@url }%
\providecommand \@url [1]{\endgroup\@href {#1}{\urlprefix }}%
\providecommand \urlprefix  [0]{URL }%
\providecommand \Eprint [0]{\href }%
\providecommand \doibase [0]{http://dx.doi.org/}%
\providecommand \selectlanguage [0]{\@gobble}%
\providecommand \bibinfo  [0]{\@secondoftwo}%
\providecommand \bibfield  [0]{\@secondoftwo}%
\providecommand \translation [1]{[#1]}%
\providecommand \BibitemOpen [0]{}%
\providecommand \bibitemStop [0]{}%
\providecommand \bibitemNoStop [0]{.\EOS\space}%
\providecommand \EOS [0]{\spacefactor3000\relax}%
\providecommand \BibitemShut  [1]{\csname bibitem#1\endcsname}%
\let\auto@bib@innerbib\@empty
\bibitem [{\citenamefont {Paul}(1990)}]{Paul:1990}%
  \BibitemOpen
  \bibfield  {author} {\bibinfo {author} {\bibfnamefont {W.}~\bibnamefont
  {Paul}},\ }\href@noop {} {\bibfield  {journal} {\bibinfo  {journal} {Rev.
  Mod. Phys.}\ }\textbf {\bibinfo {volume} {62}},\ \bibinfo {pages} {531}
  (\bibinfo {year} {1990})}\BibitemShut {NoStop}%
\bibitem [{\citenamefont {Winter}\ and\ \citenamefont
  {Ortjohann}(1991)}]{Winter:1991}%
  \BibitemOpen
  \bibfield  {author} {\bibinfo {author} {\bibfnamefont {H.}~\bibnamefont
  {Winter}}\ and\ \bibinfo {author} {\bibfnamefont {H.}~\bibnamefont
  {Ortjohann}},\ }\href@noop {} {\bibfield  {journal} {\bibinfo  {journal} {Am.
  J. Phys.}\ }\textbf {\bibinfo {volume} {59}},\ \bibinfo {pages} {807}
  (\bibinfo {year} {1991})}\BibitemShut {NoStop}%
\bibitem [{\citenamefont {Leibfried}\ \emph {et~al.}(2003)\citenamefont
  {Leibfried}, \citenamefont {Blatt}, \citenamefont {Monroe},\ and\
  \citenamefont {Wineland}}]{Leibfried:2003}%
  \BibitemOpen
  \bibfield  {author} {\bibinfo {author} {\bibfnamefont {D.}~\bibnamefont
  {Leibfried}}, \bibinfo {author} {\bibfnamefont {R.}~\bibnamefont {Blatt}},
  \bibinfo {author} {\bibfnamefont {C.}~\bibnamefont {Monroe}}, \ and\ \bibinfo
  {author} {\bibfnamefont {D.}~\bibnamefont {Wineland}},\ }\href@noop {}
  {\bibfield  {journal} {\bibinfo  {journal} {Rev. Mod. Phys.}\ }\textbf
  {\bibinfo {volume} {75}},\ \bibinfo {pages} {281} (\bibinfo {year}
  {2003})}\BibitemShut {NoStop}%
\bibitem [{\citenamefont {Haeffner}, \citenamefont {Roos},\ and\ \citenamefont
  {Blatt}(2008)}]{Haeffner:2008}%
  \BibitemOpen
  \bibfield  {author} {\bibinfo {author} {\bibfnamefont {H.}~\bibnamefont
  {Haeffner}}, \bibinfo {author} {\bibfnamefont {C.}~\bibnamefont {Roos}}, \
  and\ \bibinfo {author} {\bibfnamefont {R.}~\bibnamefont {Blatt}},\
  }\href@noop {} {\bibfield  {journal} {\bibinfo  {journal} {Phys. Rep.}\
  }\textbf {\bibinfo {volume} {469}},\ \bibinfo {pages} {155} (\bibinfo {year}
  {2008})}\BibitemShut {NoStop}%
\bibitem [{\citenamefont {Monroe}\ \emph {et~al.}(1995)\citenamefont {Monroe},
  \citenamefont {Meekhof}, \citenamefont {King}, \citenamefont {Jefferts},
  \citenamefont {Itano}, \citenamefont {Wineland},\ and\ \citenamefont
  {Gould}}]{Monroe:1995}%
  \BibitemOpen
  \bibfield  {author} {\bibinfo {author} {\bibfnamefont {C.}~\bibnamefont
  {Monroe}}, \bibinfo {author} {\bibfnamefont {D.}~\bibnamefont {Meekhof}},
  \bibinfo {author} {\bibfnamefont {B.}~\bibnamefont {King}}, \bibinfo {author}
  {\bibfnamefont {S.}~\bibnamefont {Jefferts}}, \bibinfo {author}
  {\bibfnamefont {W.}~\bibnamefont {Itano}}, \bibinfo {author} {\bibfnamefont
  {D.}~\bibnamefont {Wineland}}, \ and\ \bibinfo {author} {\bibfnamefont
  {P.}~\bibnamefont {Gould}},\ }\href@noop {} {\bibfield  {journal} {\bibinfo
  {journal} {Phys. Rev. Lett.}\ }\textbf {\bibinfo {volume} {75}},\ \bibinfo
  {pages} {4011} (\bibinfo {year} {1995})}\BibitemShut {NoStop}%
\bibitem [{\citenamefont {Jones}\ and\ \citenamefont
  {Anderson}(2000)}]{Jones:2000}%
  \BibitemOpen
  \bibfield  {author} {\bibinfo {author} {\bibfnamefont {R.}~\bibnamefont
  {Jones}}\ and\ \bibinfo {author} {\bibfnamefont {S.}~\bibnamefont
  {Anderson}},\ }\href@noop {} {\bibfield  {journal} {\bibinfo  {journal} {Rev.
  Sci. Instrum.}\ }\textbf {\bibinfo {volume} {71}},\ \bibinfo {pages} {4335}
  (\bibinfo {year} {2000})}\BibitemShut {NoStop}%
\bibitem [{\citenamefont {Macalpine}\ and\ \citenamefont
  {Schildknecht}(1959)}]{Macalpine:1959}%
  \BibitemOpen
  \bibfield  {author} {\bibinfo {author} {\bibfnamefont {W.}~\bibnamefont
  {Macalpine}}\ and\ \bibinfo {author} {\bibfnamefont {R.}~\bibnamefont
  {Schildknecht}},\ }\href@noop {} {\bibfield  {journal} {\bibinfo  {journal}
  {Proc. IRE}\ }\textbf {\bibinfo {volume} {47}},\ \bibinfo {pages} {2099}
  (\bibinfo {year} {1959})}\BibitemShut {NoStop}%
\bibitem [{\citenamefont {Siverns}\ \emph {et~al.}(2012)\citenamefont
  {Siverns}, \citenamefont {Simkins}, \citenamefont {Weidt},\ and\
  \citenamefont {Hensinger}}]{Siverns:2012}%
  \BibitemOpen
  \bibfield  {author} {\bibinfo {author} {\bibfnamefont {J.}~\bibnamefont
  {Siverns}}, \bibinfo {author} {\bibfnamefont {L.}~\bibnamefont {Simkins}},
  \bibinfo {author} {\bibfnamefont {S.}~\bibnamefont {Weidt}}, \ and\ \bibinfo
  {author} {\bibfnamefont {W.}~\bibnamefont {Hensinger}},\ }\href@noop {}
  {\bibfield  {journal} {\bibinfo  {journal} {Applied Physics B}\ }\textbf
  {\bibinfo {volume} {106}},\ \bibinfo {pages} {327} (\bibinfo {year}
  {2012})}\BibitemShut {NoStop}%
\bibitem [{\citenamefont {Mathur}\ and\ \citenamefont
  {O’Connor}(2006)}]{Mathur:2006}%
  \BibitemOpen
  \bibfield  {author} {\bibinfo {author} {\bibfnamefont {R.}~\bibnamefont
  {Mathur}}\ and\ \bibinfo {author} {\bibfnamefont {P.}~\bibnamefont
  {O’Connor}},\ }\href@noop {} {\bibfield  {journal} {\bibinfo  {journal}
  {Rev. Sci. Instrum.}\ }\textbf {\bibinfo {volume} {77}},\ \bibinfo {pages}
  {114101} (\bibinfo {year} {2006})}\BibitemShut {NoStop}%
\bibitem [{\citenamefont {Jau}\ \emph {et~al.}(2011)\citenamefont {Jau},
  \citenamefont {Benito}, \citenamefont {Partner},\ and\ \citenamefont
  {Schwindt}}]{Jau:2011}%
  \BibitemOpen
  \bibfield  {author} {\bibinfo {author} {\bibfnamefont {Y.}~\bibnamefont
  {Jau}}, \bibinfo {author} {\bibfnamefont {F.}~\bibnamefont {Benito}},
  \bibinfo {author} {\bibfnamefont {H.}~\bibnamefont {Partner}}, \ and\
  \bibinfo {author} {\bibfnamefont {P.}~\bibnamefont {Schwindt}},\ }\href@noop
  {} {\bibfield  {journal} {\bibinfo  {journal} {Rev. Sci. Instrum.}\ }\textbf
  {\bibinfo {volume} {82}},\ \bibinfo {pages} {023118} (\bibinfo {year}
  {2011})}\BibitemShut {NoStop}%
\bibitem [{\citenamefont {Noriega}\ \emph {et~al.}(2016)\citenamefont
  {Noriega}, \citenamefont {Garcia-Delgado}, \citenamefont {Gomez-Fuentes},\
  and\ \citenamefont {Garcia-Juarez}}]{Noriega:2016}%
  \BibitemOpen
  \bibfield  {author} {\bibinfo {author} {\bibfnamefont {J.}~\bibnamefont
  {Noriega}}, \bibinfo {author} {\bibfnamefont {L.}~\bibnamefont
  {Garcia-Delgado}}, \bibinfo {author} {\bibfnamefont {R.}~\bibnamefont
  {Gomez-Fuentes}}, \ and\ \bibinfo {author} {\bibfnamefont {A.}~\bibnamefont
  {Garcia-Juarez}},\ }\href@noop {} {\bibfield  {journal} {\bibinfo  {journal}
  {Rev. Sci. Instrum.}\ }\textbf {\bibinfo {volume} {87}},\ \bibinfo {pages}
  {094704} (\bibinfo {year} {2016})}\BibitemShut {NoStop}%
\bibitem [{\citenamefont {Berkeland}\ \emph {et~al.}(1998)\citenamefont
  {Berkeland}, \citenamefont {Miller}, \citenamefont {Bergquist}, \citenamefont
  {Itano},\ and\ \citenamefont {Wineland}}]{Berkeland:1998}%
  \BibitemOpen
  \bibfield  {author} {\bibinfo {author} {\bibfnamefont {D.~J.}\ \bibnamefont
  {Berkeland}}, \bibinfo {author} {\bibfnamefont {J.}~\bibnamefont {Miller}},
  \bibinfo {author} {\bibfnamefont {J.}~\bibnamefont {Bergquist}}, \bibinfo
  {author} {\bibfnamefont {W.}~\bibnamefont {Itano}}, \ and\ \bibinfo {author}
  {\bibfnamefont {D.}~\bibnamefont {Wineland}},\ }\href@noop {} {\bibfield
  {journal} {\bibinfo  {journal} {J. Appl. Phys.}\ }\textbf {\bibinfo {volume}
  {83}},\ \bibinfo {pages} {5025} (\bibinfo {year} {1998})}\BibitemShut
  {NoStop}%
\bibitem [{\citenamefont {Sias}\ and\ \citenamefont
  {K{\"o}hl}(2014)}]{Sias2014}%
  \BibitemOpen
  \bibfield  {author} {\bibinfo {author} {\bibfnamefont {C.}~\bibnamefont
  {Sias}}\ and\ \bibinfo {author} {\bibfnamefont {M.}~\bibnamefont
  {K{\"o}hl}},\ }in\ \href@noop {} {\emph {\bibinfo {booktitle} {Quantum gas
  experiments - exploring manybody states}}},\ \bibinfo {editor} {edited by\
  \bibinfo {editor} {\bibfnamefont {P.}~\bibnamefont {T{\"o}rm{\"a}}}\ and\
  \bibinfo {editor} {\bibfnamefont {K.}~\bibnamefont {Sengstock}}}\ (\bibinfo
  {publisher} {Imperial College Press},\ \bibinfo {year} {2014})\
  Chap.~\bibinfo {chapter} {12}\BibitemShut {NoStop}%
\bibitem [{\citenamefont {Johnson}\ \emph {et~al.}(2016)\citenamefont
  {Johnson}, \citenamefont {Wong-Campos}, \citenamefont {Restelli},
  \citenamefont {Landsman}, \citenamefont {Neyenhuis}, \citenamefont
  {Mizrahi},\ and\ \citenamefont {Monroe}}]{Johnson:2016}%
  \BibitemOpen
  \bibfield  {author} {\bibinfo {author} {\bibfnamefont {K.}~\bibnamefont
  {Johnson}}, \bibinfo {author} {\bibfnamefont {J.}~\bibnamefont
  {Wong-Campos}}, \bibinfo {author} {\bibfnamefont {A.}~\bibnamefont
  {Restelli}}, \bibinfo {author} {\bibfnamefont {K.}~\bibnamefont {Landsman}},
  \bibinfo {author} {\bibfnamefont {B.}~\bibnamefont {Neyenhuis}}, \bibinfo
  {author} {\bibfnamefont {J.}~\bibnamefont {Mizrahi}}, \ and\ \bibinfo
  {author} {\bibfnamefont {C.}~\bibnamefont {Monroe}},\ }\href@noop {}
  {\bibfield  {journal} {\bibinfo  {journal} {Rev. Sci. Instrum.}\ }\textbf
  {\bibinfo {volume} {87}},\ \bibinfo {pages} {053110} (\bibinfo {year}
  {2016})}\BibitemShut {NoStop}%
\bibitem [{\citenamefont {Schneider}\ \emph {et~al.}(2010)\citenamefont
  {Schneider}, \citenamefont {Enderlein}, \citenamefont {Huber},\ and\
  \citenamefont {Schaetz}}]{Schneider:2010}%
  \BibitemOpen
  \bibfield  {author} {\bibinfo {author} {\bibfnamefont {C.}~\bibnamefont
  {Schneider}}, \bibinfo {author} {\bibfnamefont {M.}~\bibnamefont
  {Enderlein}}, \bibinfo {author} {\bibfnamefont {T.}~\bibnamefont {Huber}}, \
  and\ \bibinfo {author} {\bibfnamefont {T.}~\bibnamefont {Schaetz}},\
  }\href@noop {} {\bibfield  {journal} {\bibinfo  {journal} {Nature Photon.}\
  }\textbf {\bibinfo {volume} {4}},\ \bibinfo {pages} {772} (\bibinfo {year}
  {2010})}\BibitemShut {NoStop}%
\bibitem [{\citenamefont {Gandolfi}\ \emph {et~al.}(2012)\citenamefont
  {Gandolfi}, \citenamefont {Niedermayr}, \citenamefont {Kumph}, \citenamefont
  {Brownnutt},\ and\ \citenamefont {Blatt}}]{Gandolfi:2012}%
  \BibitemOpen
  \bibfield  {author} {\bibinfo {author} {\bibfnamefont {D.}~\bibnamefont
  {Gandolfi}}, \bibinfo {author} {\bibfnamefont {M.}~\bibnamefont
  {Niedermayr}}, \bibinfo {author} {\bibfnamefont {M.}~\bibnamefont {Kumph}},
  \bibinfo {author} {\bibfnamefont {M.}~\bibnamefont {Brownnutt}}, \ and\
  \bibinfo {author} {\bibfnamefont {R.}~\bibnamefont {Blatt}},\ }\href@noop {}
  {\bibfield  {journal} {\bibinfo  {journal} {Rev. Sci. Instrum.}\ }\textbf
  {\bibinfo {volume} {83}},\ \bibinfo {pages} {084705} (\bibinfo {year}
  {2012})}\BibitemShut {NoStop}%
\bibitem [{\citenamefont {Wester}(2009)}]{Wester:2009}%
  \BibitemOpen
  \bibfield  {author} {\bibinfo {author} {\bibfnamefont {R.}~\bibnamefont
  {Wester}},\ }\href@noop {} {\bibfield  {journal} {\bibinfo  {journal} {J.
  Phys. B: At. Mol. Opt. Phys.}\ }\textbf {\bibinfo {volume} {42}},\ \bibinfo
  {pages} {154001} (\bibinfo {year} {2009})}\BibitemShut {NoStop}%
\end{thebibliography}%

\end{document}